\documentclass[compsoc, conference, a4paper, 10pt, times]{IEEEtran}

\usepackage{cite}

\usepackage{dirtytalk} 
\usepackage{balance}

\usepackage{fancyvrb} 

\usepackage{colortbl}

\usepackage{hyperref}

\usepackage{caption}
\hypersetup{
	colorlinks=true,
	linkcolor=black,
	filecolor=black,
	urlcolor=black,
	citecolor=black
}

\usepackage[a4paper, margin=1in]{geometry}
\usepackage{booktabs}
\usepackage{amssymb}
\usepackage{pifont}
\newcommand{\xmark}{\ding{55}}%

\usepackage{amsmath,amssymb,amsfonts}
\usepackage{listings}
\usepackage{color}

\definecolor{dkgreen}{rgb}{0,0.6,0}
\definecolor{gray}{rgb}{0.5,0.5,0.5}
\definecolor{mauve}{rgb}{0.58,0,0.82}

\usepackage{algorithmic}
\usepackage{graphicx}
\usepackage{textcomp}
\usepackage{xcolor}
\def\BibTeX{{\rm B\kern-.05em{\sc i\kern-.025em b}\kern-.08em
    T\kern-.1667em\lower.7ex\hbox{E}\kern-.125emX}}
\begin{document}


\title{AI-Driven Cyber Threat Intelligence Automation
}

\author{	
	\IEEEauthorblockN{
				Shrit Shah, 
                Fatemeh Khoda Parast\\
				}
		        Shah, khodapaf@uoguelph.ca\\
	
	\IEEEauthorblockA{
           			Faculty of Computer Science,
			University of Guelph, Canada
		}
}

\maketitle

\begin{abstract}

This study introduces an innovative approach to automating Cyber Threat Intelligence (CTI) processes in industrial environments by leveraging Microsoft's AI-powered security technologies. Historically, CTI has heavily relied on manual methods for collecting, analyzing, and interpreting data from various sources such as threat feeds. This study introduces an innovative approach to automating CTI processes in industrial environments by leveraging Microsoft's AI-powered security technologies. Historically, CTI has heavily relied on manual methods for collecting, analyzing, and interpreting data from various sources such as threat feeds, security logs, and dark web forums—a process prone to inefficiencies, especially when rapid information dissemination is critical. By employing the capabilities of GPT-4o and advanced one-shot fine-tuning techniques for large language models, our research delivers a novel CTI automation solution. The outcome of the proposed architecture is a reduction in manual effort while maintaining precision in generating final CTI reports. This research highlights the transformative potential of AI-driven technologies to enhance both the speed and accuracy of CTI and reduce expert demands, offering a vital advantage in today's dynamic threat landscape.

\end{abstract}


\section{Introduction}


Cyber Threat Intelligence (CTI) collects, analyzes, and disseminates information about current and potential cyber threats. CTI involves data collection from various sources, such as security logs, threat feeds, and incident reports, to identify Indicators of Compromise (IoC) and understand the Tactics, Techniques, and Procedures (TTPs) used by threat actors~\cite{wagner2019cyber}. CTI is crucial in cybersecurity as it enables organizations to proactively defend against cyber attacks by providing actionable insights that inform security measures, enhance incident response, and facilitate identifying and mitigating vulnerabilities~\cite{shad2024ai, siracusano2023time, preuveneers2021sharing}.

Unlike measurement data, which is automatically gathered, threat intelligence requires meticulous manual analysis by expert analysts~\cite{schlette2021measuring, ramsdale2020comparative}. These professionals use their expertise and intuition to extract IoCs and attack patterns. An IoC could be any observable element linked to an attack, such as a file (identified by its name or hash), a URL, or an IP address. To decode the attack pattern of an attack group, analysts identify several IoCs related to a malicious campaign and ascertain the role of each one—for instance, determining if a URL is an exploitation site or a command-and-control server. This carefully curated threat intelligence is published in technical blogs and industry reports. Moreover, commercial or open-source IoC feeds share some of the identified indicators essential for detecting security threats, conducting forensic investigations after breaches, and attributing attacks to specific threat actors~\cite{shad2024ai, zhu2018chainsmith, siracusano2023time}.

In today's rapidly evolving cybersecurity landscape, the promptness and accuracy of CTI reports are crucial for safeguarding digital infrastructures by enabling immediate responses to security risks. However, the current methodologies for generating these reports predominantly rely on manual human efforts, which introduces significant challenges affecting their efficiency and effectiveness. The manual approach to report generation is inherently time-consuming, labour-intensive, and susceptible to human error. Threat Intelligence Analysts must sift through vast amounts of data from multiple sources, synthesize the information, and compile comprehensive reports, all of which demand substantial expertise and meticulous attention to detail~\cite{alam2023looking, husari2017ttpdrill}. 

In the cybersecurity community, several standard methods are used to automate CTI processes involving open-source tools, custom scripting, platform integration, and AI-based methods, which still demand CTI specialists~\cite{li2022attackg, jo2022vulcan, zhao2020timiner}. 
This study addresses CTI automation using AI-powered Microsoft security products, including Microsoft Copilot for Security (MCS), Logic Apps, and Azure AI, within an industrial-scale environment. 
The proposed method offers significant benefits, including faster threat detection, improved response times, and more accurate threat intelligence. Out method facilitates CTI processes and reduces manual efforts, ultimately strengthening cybersecurity defences.
The experimental results, along with interviews with security experts, have revealed a significant impact of these technologies.
We present a detailed architectural design for deploying this method within an industrial environment, highlighting how this approach can enhance security efficiency and effectiveness.

\subsection{Problem Statement}

Many organizations today rely on manual and semi-automated approaches to collect, analyze, and generate CTI reports. These methods often involve threat analysts gathering threat data from disparate sources, such as security logs, threat feeds, vulnerability databases, and social media platforms. Analysts must then filter through this data and manually correlate it to identify patterns and assess the potential impact on the organization's security posture. This process may involve using essential tools, but in many cases, it still requires a significant amount of manual work, including writing the final report\cite{schaberreiter2019quantitative}. 
The manual nature of these methods introduces several challenges. The gathering, analyzing, and reporting process can take days or weeks, delaying the response to emerging threats. Due to the complexity of modern threats, manual processing increases the likelihood of missing critical insights or introducing errors. Different analysts may interpret threat data differently, leading to inconsistencies in reporting.
As the volume and complexity of threats increase, manual processes struggle to keep pace, leading to bottlenecks in threat detection and mitigation efforts~\cite{schaberreiter2019quantitative, ramsdale2020comparative}.

\begin{figure}
	\begin{center}
		\includegraphics[width=0.35\textwidth]{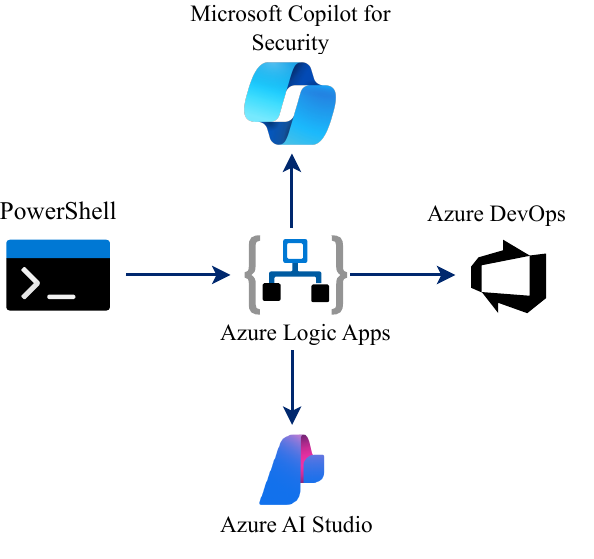}
			\caption {Automated CTI Generation Architecture}	\label{fig:flowdiagram}
	\end{center}
\end{figure}

In response to these challenges, automation and using AI offers promising solutions. Organizations can significantly reduce the time and effort required to generate final CTI reports by automating the collection and analysis of threat data. AI-powered systems can ingest vast amounts of threat data in real-time, correlate patterns across multiple sources, and identify emerging threats with high accuracy. Machine learning models can learn continuously from new data, improving their detection capabilities and reassuring organizations of their adaptability to new threats as they evolve~\cite{alam2023looking, husari2017ttpdrill}.
Despite these advancements, many organizations are hesitant to adopt fully automated CTI processes due to concerns about the reliability of AI-driven systems, particularly in complex or high-stakes environments. They may worry about false positives, the inability of machines to interpret nuanced threat information, or the lack of transparency in how AI makes decisions~\cite{jo2022vulcan}.

This research aims to address the gap by exploring the limitations of current manual CTI report generation processes and investigating how automated solutions, including AI, can enhance these reports' overall quality, speed, and accuracy. By identifying which tasks within the manual process can be effectively automated—such as data collection, correlation, and preliminary analysis—this study aims to lay the groundwork for developing more advanced automated CTI systems. These systems will not replace threat intelligence analysts. However, they will better support them by minimizing the manual effort required in report creation, thus enabling faster and more consistent dissemination of threat intelligence. Applying the proposed method, in turn, will allow organizations to respond more rapidly to emerging threats and implement defensive measures promptly.
This research offers a practical solution to the critical shortage of highly skilled cyber threat analysts worldwide, particularly impacting medium-sized cybersecurity companies that provide CTI services.

\section{Current Trends and Scholarly Perspectives}

\begin{table*}
	\begin{center}
		\caption{Summary of Current Studies on Automated CTI Extraction}
		\begin{tabular}{   
				p{0.13\textwidth}   
				>{\centering\arraybackslash}p{0.04\textwidth}   
				>{\centering\arraybackslash}p{0.05\textwidth}
				>{\centering\arraybackslash}p{0.05 \textwidth}  
				>{\centering\arraybackslash}p{0.05\textwidth}  	
				>{\centering\arraybackslash}p{0.08\textwidth}  
				>{\centering\arraybackslash}p{0.05\textwidth}   					
				p{0.3\textwidth}  } 
			\toprule
			Research & Year & IoC  & TTP  & APT & Mitigation & MITRE & Method \\
			\midrule
			\rowcolor[HTML]{EFEFEF}
			TTPDRIL~\cite{husari2017ttpdrill} & 2017 & \checkmark & \checkmark & \xmark  & \xmark          & 	\checkmark  &	 NLP, Information Retrieval	\\
			
			TIMiner~\cite{zhao2020timiner} & 2020 & \checkmark  & \checkmark &  \xmark & \xmark & \xmark &	 Convolutional Neural Network \\
			
				\rowcolor[HTML]{EFEFEF}
			Extractorç& 2021 & \checkmark &\checkmark & \xmark  & \xmark&  \xmark& NLP, Semantic Role Labeling	\\
			
				AttacKG~\cite{li2022attackg} & 2022 & \checkmark &  \checkmark & \checkmark  & \xmark &  \checkmark &	Knowledge Graph\\
			
				\rowcolor[HTML]{EFEFEF}
			TIM~\cite{you2022tim}& 2022 &  \checkmark & \checkmark  &  \checkmark  & \xmark& \checkmark & Deep Learning Models,  Classification	\\
			
			TriCTI~\cite{liu2022tricti} &  2022 & \checkmark &\checkmark  & \checkmark  & \checkmark & \xmark & NLP, Data augmentation\\
			
			\rowcolor[HTML]{EFEFEF}
			Vulcan~\cite{jo2022vulcan} & 2022 & \checkmark & \checkmark &  \checkmark & \xmark  & \xmark & LLM, Named Entity Recognition	\\
			
			LADDER~\cite{alam2023looking} & 2023&\checkmark&  \checkmark & \checkmark & \xmark & \checkmark & Knowledge Graph \\
		
			\bottomrule
		\end{tabular}
	\end{center}
\end{table*}

	Numerous studies have proposed automating the final extraction of CTI using AI-based methods, including Natural Language Processing (NLP),  Information Retrieval (IR), and Machine Learning (ML) approaches such as Neural Networks (NN)~\cite{husari2017ttpdrill, zhao2020timiner, satvat2021extractor, jo2022vulcan}. Husari et al.~\cite{husari2017ttpdrill} proposed the pioneer model TTPDrill to automatically extract threat actions and map them to attack patterns, techniques, and phases of the cyber kill chain from unstructured CTI reports. The tool combines NLP and IR techniques, using an ontology derived from MITRE ATT\&CK and CAPEC repositories---Common Attack Pattern Enumeration and Classification or CAPEC~\cite{CAPEC}, is a comprehensive database maintained by MITRE Corporation---to map extracted TTPs to recognized cyber threat patterns. The repository aims to provide a structured and detailed understanding of how attacks are executed, which helps cybersecurity professionals anticipate threats, design effective defences, and communicate about security issues more effectively. TTPDrill aims to facilitate timely, cost-effective cyber defence by converting unstructured CTI into structured threat information, significantly improving the precision and recall of TTP extraction. 

	Zhao et al.~\cite{zhao2020timiner}  presents TIMiner, an automated framework designed to extract and categorize CTI from unstructured social media data. The framework uses a Convolutional Neural Network (CNN) to classify CTIs into specific domains such as finance, government, IoT, and Industrial Control Systems (ICS). The study employs a hierarchical IoC extraction method that leverages word embedding and syntactic dependency to identify known and unknown types of IoCs. TIMiner also introduces a novel metric, Threat-Index, which quantifies the severity of threats across different domains by analyzing attack frequency, exploited vulnerabilities, and risk levels.

	Satvat et al.~\cite{satvat2021extractor} introduce an EXTRACTOR tool, which automates the extraction of concise attack behaviour from unstructured CTI reports. EXTRACTOR generates provenance graphs that capture the relationships between entities, actions, and system artifacts in a cyber attack. The tool leverages NLP techniques like Semantic Role Labeling (SRL) to process complex CTI texts and infer \textit{who did what to whom} in an attack. The study addresses verbosity, syntactic complexity, and the need to capture causal and temporal dependencies between actions. EXTRACTOR generates actionable intelligence, simplifying threat-hunting tasks by focusing on attack behaviour that can be observed in system audit logs. The tool was evaluated against real-world CTI reports and DARPA adversarial engagements, showing high precision and utility in detecting cyber threats using automatically generated graphs.
	
	Jo et al.~\cite{jo2022vulcan} present a CTI system called Vulcan, which automates the extraction of diverse CTI data, including TTPs, from unstructured sources like security reports and blogs. Vulcan employs a fine-tuned language model to perform Named Entity Recognition (NER) and Relation Extraction (RE), accurately identifying CTI entities (e.g., attack vectors, tools) and their relationships. The extracted CTI data are stored in a graph database, accessible via APIs, enabling applications such as threat evolution tracking and profiling. Liu et al.~\cite{liu2022tricti} introduce a system to extract actionable CTI from unstructured cybersecurity reports automatically. TriCTI uses a trigger---enhanced neural network model that incorporates \textit{campaign triggers}---to classify IoCs based on the cyber kill chain stages. By leveraging NLP and data augmentation techniques, TriCTI improves classification accuracy and generates actionable intelligence that can be integrated into intrusion detection systems.

	You et al.~\cite{you2022tim} introduce the TIM framework designed to mine and classify TTPs from unstructured threat data such as security analysis reports. The framework uses a TCENet (Threat Context Enhanced Network) model that leverages textual descriptions and specific TTP-related elements (e.g., IPs, CVEs, file hashes) to classify TTP. The lack of available sentence-level TTP datasets and the complexity of accurately classifying TTP descriptions is addressed in this study. In a recent study, Le et al.~\cite{alam2023looking} introduced LADDER, a framework designed for knowledge extraction that identifies text-based attack patterns from CTI reports. LADDER automates the extraction of TTPs and other essential data related to malware and APTs from CTI sources. The framework organizes the extracted information using an ontology—a structured representation of knowledge domains in information science—to aid in meaningful information organization. Combined with the TTPClassifier, this ontology integrates the information into a comprehensive knowledge graph (KG). The KG, formatted as triples (subject, predicate, object), includes domain-specific information, facilitating predictive analysis. In contrast to our research, the author proposed a method to automate only attack pattern extraction. Our work not only automates the whole CTI generation but also produces a comprehensive report, including strategic and operational CTI. 
	
	Li et al.~\cite{li2022attackg} introduce AttacKG, a novel system designed to automatically extract structured attack behaviour graphs from unstructured CTI reports and aggregate them to form a Technique Knowledge Graph (TKG). This approach addresses the challenge of manually recovering attack behaviours from CTI reports, often written in natural language. By parsing multiple CTI reports, AttacKG identifies and integrates diverse attack techniques, creating comprehensive knowledge graphs that summarize various aspects of attack patterns. The system outperforms existing tools in accuracy, providing valuable intelligence for security tasks like APT detection and cyber attack reconstruction.

	Recent advancements in AI-driven CTI extraction, such as LADDER and AttacKG, employ a combination of NLP techniques and open-source pre-trained LLMs, including BERT~\cite{devlin2018bert}, RoBERTa~\cite{liu2019roberta}, and XLM-RoBERTa~\cite{conneau2019unsupervised}, to identify relevant sentences, determine if they describe an attack, and map these attacks to the corresponding Tactics, Techniques, and Procedures (TTPs) in the MITRE ATT\&CK framework~\cite{jo2022vulcan, zhao2020timiner, shad2024ai, siracusano2023time, zhu2018chainsmith}. Despite their effectiveness, these approaches still demand specialized expertise for model implementation, limiting their practicality in environments without such resources. 	
	In contrast, our study introduces the first fully automated CTI extraction methodology that eliminates the need for domain-specific expertise. Additionally, our approach autonomously generates comprehensive CTI reports, covering IoCs and TTPs, mapping attack behaviours to the MITRE framework, and recommending mitigation strategies. These reports are organized into two primary sections: strategic reports, offering high-level summaries for executives, and technical reports, delivering detailed operational intelligence.

\section{Methodology}

	This study introduces a fully automated, end-to-end process for generating CTI reports using AI built within the Microsoft enterprise ecosystem. Key technologies include PowerShell scripting, Azure Logic Apps, MCS, Azure AI Studio, and Azure DevOps. The automation begins with a PowerShell script, which gathers essential data from the user and manages the workflow. Azure Logic Apps then orchestrate the report generation by dividing it into multiple sections. Each section is produced through customized prompts directed to MCS  or Azure AI. Once all sections are completed, the Logic Apps and PowerShell script merges them into a final, cohesive report that is automatically stored in the Azure DevOps repository. The proposed architecture is shown in Figure~\ref{fig:flowdiagram}. The segmentation approach ensures precise control over the prompt creation, allowing for customization of each section. 
	The generated final CTI report comprises the following seven sections:
	
\begin{enumerate}  
	\item Metadata and Overview  
	\item MITRE Summary Table  
	\item Data Extraction  
	\item Tools and Malware  
	\item Defense Recommendations  
	\item References  
	\item Tags  
\end{enumerate}  

The first and second sections include strategic CTI information for executive-level decision-making, while the remaining sections fall under the technical CTI report category.
The \textit{Metadata and Overview} section provides general information about the report, including the report title, creation date, attacker details, and a threat intelligence summary. The \textit{Overview} briefly describes the scope of the threat, including the affected sectors, systems, or technologies, and the severity or priority level of the threat, offering a high-level snapshot to set the context for the rest of the report. 
The \textit{MITRE Summary Table} section presents a structured summary of the identified TTPs used in the threat, mapped to the MITRE ATT\&CK framework, including specific techniques leveraged by attackers and highlighting the associated adversarial tactics. The table serves as a quick reference for understanding the threat actors' behaviour and operational methodologies.

The \textit{Data Extraction} section contains all relevant data points related to the observed threat. This section typically includes IoCs information, such as IP addresses, file hashes, domain names, and other artifacts that may assist in identifying or tracking a threat. This section focuses on the factual data collected from various sources during the investigation. 
The \textit{Tools and Malware} section details any tools, exploits, or malware observed in the attack, including the descriptions of how these tools are used by threat actors and their potential impact on compromised systems. The analysis often highlights known or new malware variants, software used in the attack chain, and the relationships between the tools and broader campaigns.

The \textit{Defense Recommendations} section provides actionable recommendations to mitigate or respond to threats. This section covers suggested defensive measures, including patches, configurations, detection rules, and long-term strategies to protect against similar future threats. It may also highlight gaps in current defences that organizations should address. 
The \textit{References} section lists all the sources used to create the CTI report, including but not limited to academic papers, threat feeds, public and private security advisories, and other research material relevant to the threat. 
Finally, the \textit{Tags} section contains keywords or phrases that categorize the threat for easier indexing and retrieval. Tags may include the malware family name, associated adversaries, affected industries, geographic regions, TTPs, or specific vulnerabilities exploited. These tags help in searching and cross-referencing the report with other similar threats.

\subsection{PowerShell Script}  \label{sec:powershell}

\begin{figure}
	\begin{center}
		\includegraphics[width=0.46
		\textwidth]{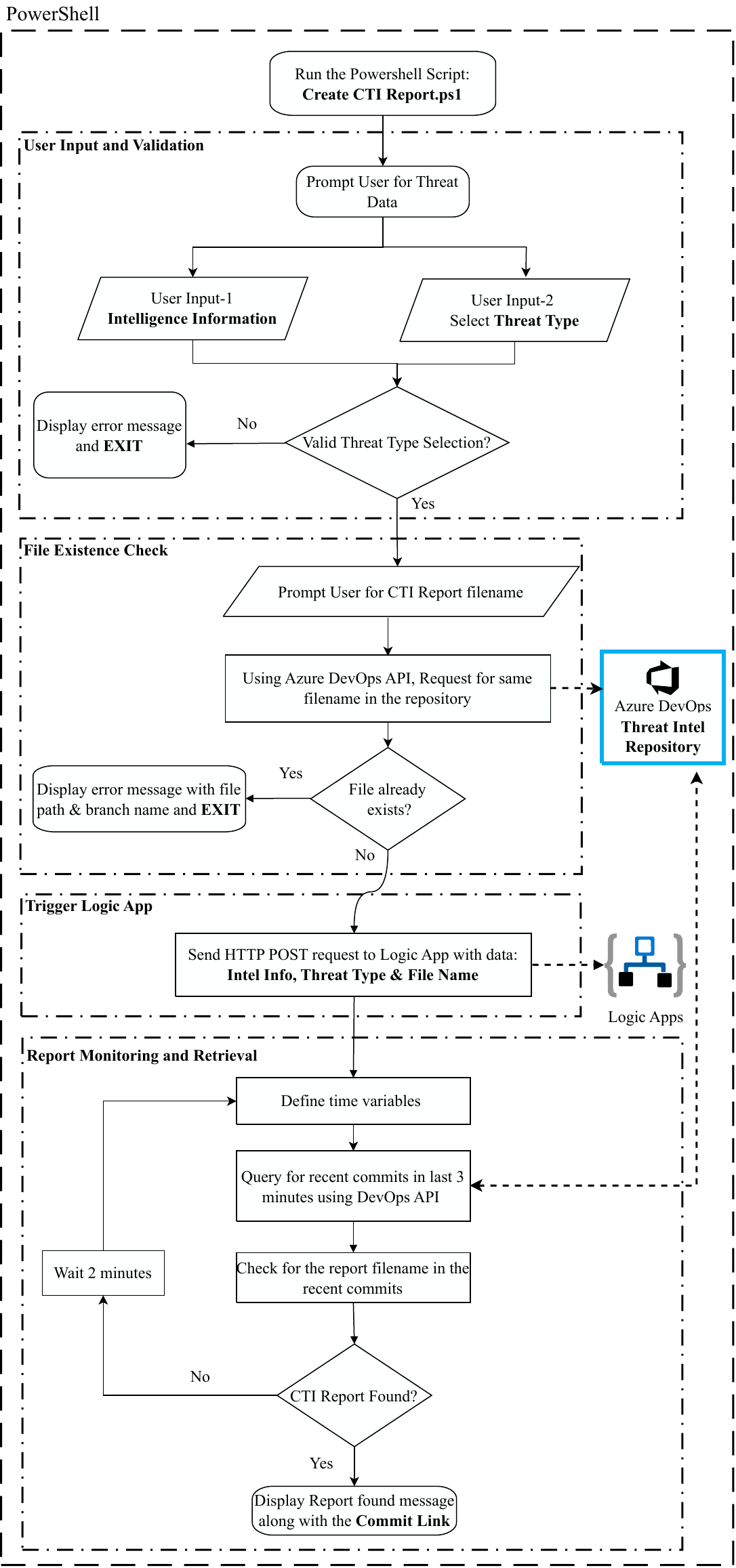}
			\caption {PowerShell Architecture Diagram}	\label{fig:PowerShell}
	\end{center}
\end{figure}

The PowerShell script acts as the main user interface, orchestrating the initialization of all the underlying systems, including Azure Logic Apps and Azure DevOps. This component integrates various functionalities such as user input, file validation, HTTP requests, and monitoring mechanisms to ensure a smooth and automated workflow. As illustrated in Figure~\ref{fig:PowerShell}, the script's tasks are divided into four main functions:

\begin{enumerate}  
	\item User Input and Validation
	\item File Existence Check
	\item Trigger Logic Apps
	\item Report Monitoring and Retrieval
\end{enumerate}

Based on artifact similarities, the threat types are categorized as follows: Campaign, Threat Actor, Vulnerability, or Malware/Tool.
The \textit{user input} and \textit{validation} functions collect intelligence data and verify the threat type.
The script then prompts the user to provide essential intelligence information 
and select a specific threat type, ensuring the chosen threat meets predefined criteria for validity. 
The \textit{file existence check} ensures the uniqueness of the CTI report's file name. To start, the user is prompted to input a file name checked against the Azure DevOps repository to avoid duplicates. If a file with the same name already exists, the user is notified and asked to provide a new name.
The \textit{Trigger Logic Apps} function initiates the CTI report creation process by collecting all user-provided data and sending an HTTP request to the Logic Apps. This data is passed as a JSON request body, and the function ensures the request is successfully sent while handling potential errors.
The \textit{report monitoring} and \textit{retrieval} components continuously monitor the Azure DevOps repository for the CTI report's creation. The script checks for recent commits every two minutes until the CTI report is located. Once the report is successfully generated, the user is notified that it is available in the repository.

\subsection{Azure Logic Apps}  

	Azure Logic Apps defines a cloud service that helps users automate workflows and integrate applications, data, and services across organizations. It allows the creation of automated workflows to connect different services and applications and offers numerous pre-built connectors for services. Users can create custom workflows using a visual designer or code and trigger workflows based on specific events or schedules. Azure Logic Apps automates business processes and tasks, integrates disparate systems and data sources, manages enterprise workflows and data flows, and builds complex integration scenarios with minimal coding.

\begin{figure}
		\centering
	\includegraphics[width=0.45\textwidth]{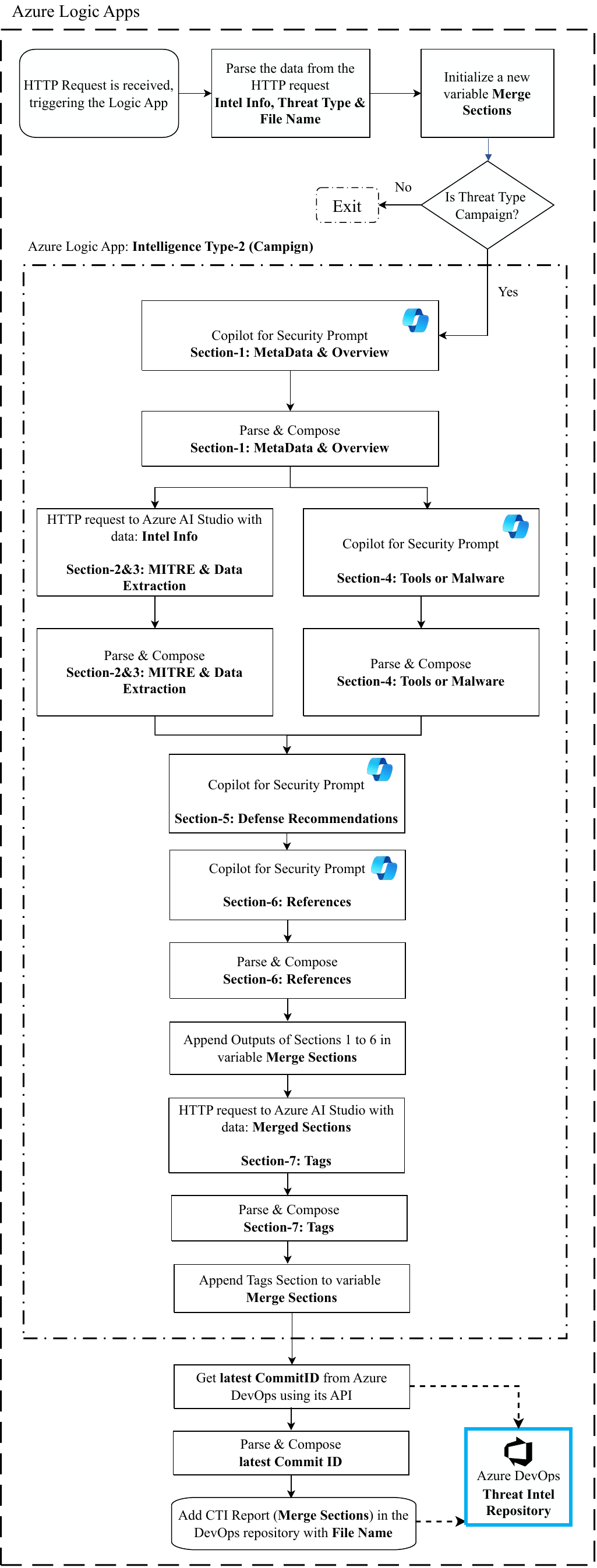}
	\caption{Azure Logic Apps Architecture Diagram} 	\label{fig:LogicApp}
\end{figure}

The Azure Logic Apps workflow serves as the core engine of the process, orchestrating the generation of the CTI report by integrating MCS, Azure AI Studio, and Azure DevOps—a Cloud-based platform from Microsoft that offers tools for managing the entire software development lifecycle, including version control, project tracking, and collaboration. This workflow is designed to manage various types of threat intelligence, including campaigns, threat actors, vulnerabilities, and malware/tools. Depending on the threat type, the workflow branches into sub-workflows, employing Azure AI and MCS where applicable. The process is initiated by an HTTP request from a PowerShell script, which sends JSON-formatted data containing key parameters such as intelligence information, threat type, and file name. A variable, \textit{Merge Sections}, is created to store the context of the report sections. The Logic Apps workflow then assesses the threat type (Campaign, Threat Actor, Vulnerability, or Malware/Tool) and redirects the process accordingly, as illustrated in Figure~\ref{fig:LogicApp}. The figure provides the \textit{Campaign} threat type as an example for clarity without loss of generality.

	The CTI report is organized into distinct sections based on the threat type (elaborated in Section~\ref{sec:powershell}). Logic Apps initiates Azure AI or MCS to generate each section accordingly, appending the results to the \textit{Merge Sections} variable. Once all sections are compiled, Logic Apps retrieves the latest commit details from the Azure DevOps repository. The response is parsed to extract the most recent commit ID, which is required to initiate a new commit via the DevOps API. Finally, the fully compiled CTI report, stored within the \textit{Merge Sections} variable, is uploaded to the Azure DevOps repository under the specified file name using the DevOps API.

\begin{figure*}
	\centering
	\includegraphics[width=1\textwidth]{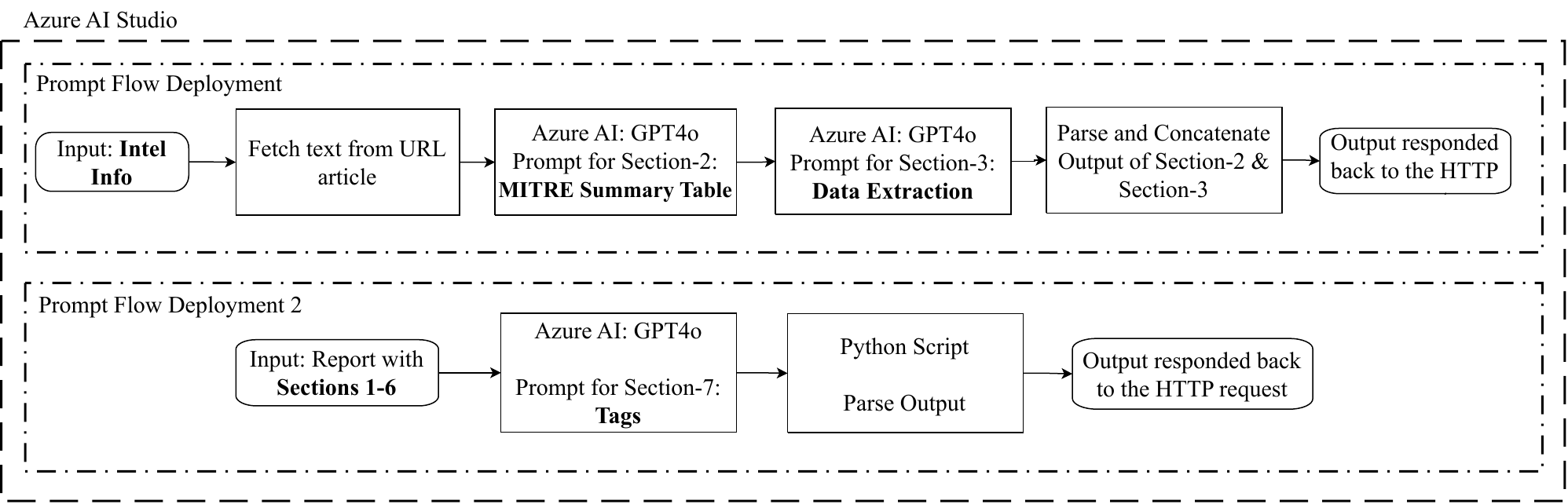}
	\caption{Azure AI Studio Architecture Diagram}	\label{fig:AzureAI}
\end{figure*}

\subsection{Azure AI Studio}  

Azure AI Studio facilitates the development of prompt flows, empowering developers to design, debug, and deploy AI-driven applications using customizable prompts and Python scripts. In this architecture, these flows are integrated with Azure Logic Apps. Our design leverages OpenAI's GPT-4o model to extract, analyze, and generate specific sections of the CTI report. As depicted in Figure~\ref{fig:AzureAI}, the workflow is structured into two main prompt flow deployments: one dedicated to generating the MITRE summary table and performing data extraction, and the other focused on generating the Tags section.
In the first prompt flow, a spreadsheet containing the complete MITRE ATT\&CK enterprise matrix is uploaded into the Azure AI environment to facilitate prompt flow deployment. This file provides a comprehensive reference for ATT\&CK techniques and sub-techniques. The process initiates when intelligence information (Intel Info), received through an HTTP request from Logic Apps, is sent to Azure AI Studio. A Python script is subsequently employed to extract relevant text from a specified URL or article, which is the basis for further analysis and data extraction.

The retrieved text is processed by prompting GPT-4o to generate a summary table of MITRE ATT\&CK techniques. In this step, GPT-4o is tasked with creating a section explaining the techniques and sub-techniques relevant to the specific campaign.
Subsequently, a Python script parses the outputs from the earlier steps and merges the results of Section 2 (MITRE Summary Table) and Section 3 (Data Extraction). After the first prompt flow, the combined output is returned as an HTTP response to the corresponding Logic Apps action.
The second prompt flow deployment handles the generation of tags for CTI reports based on their content. The process begins with the input of a report containing Sections 1-6 into the Azure AI Studio. Then Azure AI prompts GPT-4o to generate tags for Section-7 of the CTI report, based on the content from Sections 1-6. At the end of the second flow, a Python script parses the output generated by Azure AI.

\subsection{Microsoft Copilot for Security}

	\textbf{MCS} is an AI-powered assistant designed to enhance cybersecurity operations by providing real-time insights and recommendations. MCS, powered by machine learning, efficiently analyzes threat data and provides actionable insights, helping security teams respond to threats more effectively with guided steps and automation. MSC assists proactive threat hunting by analyzing vast amounts of security data to integrate the outcome with existing Microsoft security tools such as Microsoft Defender and Sentinel.  
In the proposed architecture, MCS is integrated with Azure Logic Apps, which provide a built-in action to facilitate direct interaction with MCS, enabling prompts to be sent and responses to be received within the workflow. Additionally, Azure Logic Apps allows the specification of \textit{skills}, which are capabilities that enhance prompt processing. These skills enable extended functionality, such as accessing web URLs or retrieving threat intelligence from Microsoft Defender for threat intelligence. In this architecture, MCS generates four essential sections of the report: Metadata and Overview, Tools or Malware, Defense Recommendations, and References.
From this point on, we use AI, LLM, and GPT-4.0 interchangeably to refer to systems that receive prompts and generate outputs. Specific distinctions will be made when necessary.


\subsubsection{Metadata and Overview}

At this stage, the prompt directs the AI to extract detailed information about a threat campaign from a URL. Then, it is tasked with populating a pre-defined CTI report template in markdown format, ensuring accuracy and thoroughness. The critical elements, such as the campaign name, threat type, associated threat groups, and targeted industries, are filled in based on the data retrieved from the source. Additionally, the AI generates a strategic campaign summary to enhance accessibility for a wider audience. This structured process ensures the creation of consistent, detailed CTI reports while minimizing the risk of human error.

\subsubsection{Tools or Malware}

To generate this section, the prompt concentrating on identifying malicious software, tools, or malware employed in a cyber threat campaign. The AI generates a detailed technical report using data from a provided URL and the Microsoft Defender CTI Platform. A structured markdown template captures specific sections for each identified tool or malware. For each entry, the AI delivers a description and an example of how the threat actors utilize the tool or malware. This approach ensures the report is technically comprehensive and accurate, leveraging authoritative sources for in-depth analysis.

\subsubsection{Defense Recommendations}

In this section, the prompt identifies malicious software, tools, or malware employed in a cyber threat campaign. The LLM generates a detailed technical report using data from a provided URL and the Microsoft Defender CTI Platform. A structured markdown template captures specific sections for each identified tool or malware. For each entry, the AI delivers a description and an example of how the threat actors utilize the tool or malware. This approach ensures the report is technically comprehensive and accurate, leveraging authoritative sources for in-depth analysis.

\subsubsection{References}

The  LLM first compiles a list of references specific to a given threat campaign in this prompt. It gathers data from the provided URL and other publicly available sources. A structured markdown template populates these references as a list of URLs related to the threat campaign, ensuring the accurate collection and citation of relevant sources. This process provides comprehensive references to substantiate the report's findings.

\section{Performance Assessment and Key Findings}

We evaluated the proposed approach from two perspectives: performance and cost. Initially, we conducted experiments comparing the reports generated manually by CTI analysts with those produced by AI. This phase focused on a \textit{Campaign} threat type, comparing eight attack campaigns created by both methods, which are extracted from hundreds of reports, news, blogs, and other open and closed CTI resources. Lastly, we assessed and presented the cost associated with AI-based report generation using the proposed method.

\subsection{Performance Evaluation Manual vs. AI-Based Methods}

\begin{table*}
		\caption{Comparison of BERT and Cosine Similarity}\label{tab:BERT-Cosine}
		\begin{tabular}{   
				p{0.2\textwidth}   
				p{0.41\textwidth}   
				p{0.31\textwidth}  
			} 
			\toprule
			Feature & BERT Embeddings  &  Cosine Similarity      \\
			\midrule
			\rowcolor[HTML]{EFEFEF}
			{Representation}          & Contextual, dense vector embeddings                           & Word frequency or TF-IDF vector                  \\ 
			{Context Sensitivity}     & High (captures relationships and context)                    & None (words treated independently)           \\ 
			
			\rowcolor[HTML]{EFEFEF}
			{Semantic Awareness}      & High (understands synonyms and related concepts)             & Limited (no understanding of synonyms)         \\
			{Dimensionality}          & Lower-dimensional and dense                                  & High-dimensional and sparse                       \\ 
			
			\rowcolor[HTML]{EFEFEF}
			{Computational Complexity} & High (requires pre-trained model and more processing)         & Low (faster to compute)                                \\ 
			{Sensitivity to Word Order} & Captures meaning of word order and sentence structure        & None (word order ignored)                   \\ 
			
			\rowcolor[HTML]{EFEFEF}
			{Vocabulary Specificity}   & Can generalize to similar or related words                   & Requires exact word matches                 \\ 
			{Applicability}            & Better for understanding deep semantic meaning               & Suitable for simple keyword overlap         \\ 
			\bottomrule
		\end{tabular}
	\end{table*}

We selected Cosine Similarity and BERT (Bidirectional Encoder Representations from Transformers) to compare the AI-generated report with the manual counterpart. BERT can process up to 512 tokens at a time, allowing it to handle relatively long documents~\cite{alewiwi2016efficient, srivastava2023new, guo2020detext}.
BERT  can be used to compare two documents by representing each document as a numerical vector (embedding) and then measuring the similarity between these vectors. The process involves encoding the documents' text into high-dimensional embeddings that capture the text's semantic meaning, allowing for a meaningful comparison.  
BERT starts by breaking down the text of each document into tokens (typically words) and converting words into a sequence of token IDs that BERT can process. Then, it adds unique tokens like \textit{[CLS]} at the start of each document and \textit{[SEP]} at the end to signify the boundaries of the input. Each token is then mapped to a pre-trained embedding vector using BERT's embedding layer. These embeddings are contextual, meaning that the representation of a word depends on the words around it. BERT uses multiple layers of bidirectional transformers to process the input sequence and generate contextualized embeddings for each token. The embeddings capture the meaning of the words considering the surrounding context rather than treating them as isolated entities.

After processing the document through BERT, each document is represented as a high-dimensional vector (embedding). These vectors capture the meanings of individual words and their relationships in the document's context. Once both documents have been transformed into vectors, we compare them using a Cosin Similarity. 
We also use Cosine Similarity to capture the similarity between the vocabulary of the two documents. 
The formula for cosin similarity between two vectors \( A \) and \( B \) is:

\[
\text{cosine similarity} = \frac{A \cdot B}{\|A\| \|B\|}
\]

\noindent
where \( A \cdot B \) is the dot product of the two vectors, and \( \|A\| \) and \( \|B\| \) are the magnitudes (or norms) of vectors \( A \) and \( B \), respectively. The expanded formula then becomes:

\[
\text{cosine similarity} = \frac{\sum_{i=1}^{n} A_i B_i}{\sqrt{\sum_{i=1}^{n} A_i^2} \cdot \sqrt{\sum_{i=1}^{n} B_i^2}}
\]

Cosine similarity measures the cosine of the angle between two non-zero vectors. A result of 1 indicates the vectors point in the same direction (maximum similarity), 0 means the vectors are orthogonal (no similarity), and -1 indicates the vectors point in opposite directions (maximum dissimilarity)~\cite{alewiwi2016efficient}.

As an example, comparing \textit{\say{According to a new report by Trustwave, cybercriminals have developed an innovative phishing method that involves the use of encrypted RPMSG attachments.}} with \textit{\say{The article from Trustwave discusses a phishing campaign that uses Microsoft Encrypted Restricted Permission Messages to deliver phishing attacks.}} results in a  89\% and 22\% similarity using BERT and Cosine Similarity. While BERT captures the context, the latter indicates more sensitivity to the wording. Table~\ref{tab:BERT-Cosine} represent a comparison between the two metrics.

\begin{figure}
	\includegraphics[width=0.5\textwidth]{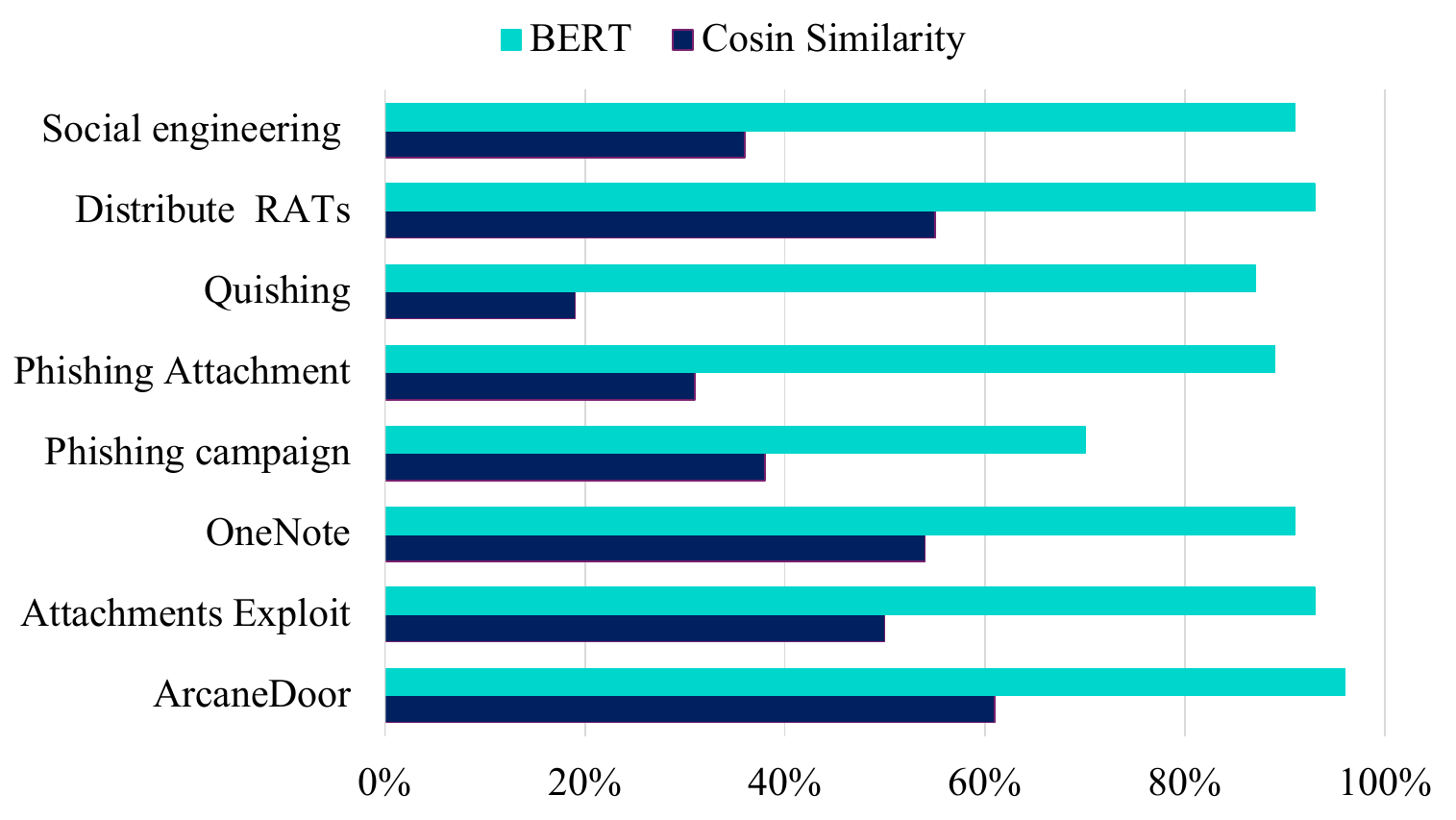}
	\caption{Similarity of AI-generated vs Manual Strategic CTI Reports.}\label{fig:BERT-vs-CS}
\end{figure}

The final CTI reports provide multiple sections to cater to different stakeholders. Two key sections are the executive (or strategic) summaries, designed to support informed decision-making at the executive level. The remaining sections offer a technical analysis, equipping the security operations team with the necessary information to implement specific security measures. Our experiments have validated the precision of each report tier. As illustrated in Figure~\ref{fig:BERT-vs-CS}, we compared the strategic-level sections of threat reports using BERT and Cosine Similarity metrics. While the average Cosine Similarity was 43\%, the BERT model achieved an 89\% score, demonstrating that the reports effectively conveyed the same context even with different terminologies.

\begin{figure}
	\includegraphics[width=0.45\textwidth]{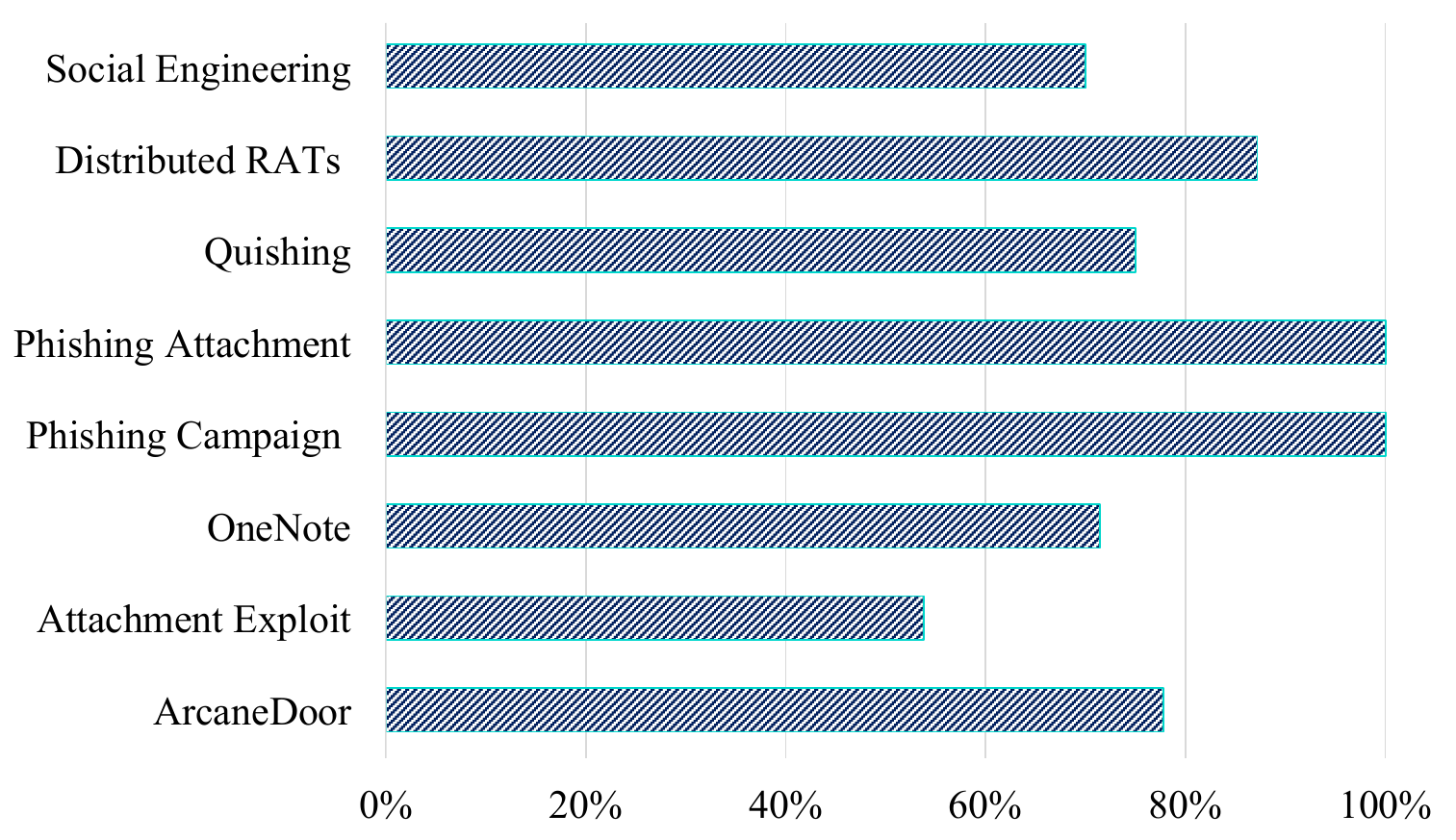}
	\caption{Identification Accuracy of AI-generated  CTI Reports Attack Pattern Extraction.} 	\label{fig:Manual-vs-AI-TTP-Extraction}
\end{figure}

The second experiment evaluates the extraction of attack patterns by comparing the manual methods versus AI models. We used  Accuracy to report the performance of the model. 

\[
\text{Accuracy} = \frac{\text{Number of Correct Predictions}}{\text{Total Number of Predictions}}
\]

\noindent
where the term \textit{Total Number of Predictions} refers to the complete set of TTPs that should have been detected, while \textit{Number of Correct Predictions} represents the TTPs successfully identified using the AI method. Figure~\ref{fig:Manual-vs-AI-TTP-Extraction} illustrates the performance of the AI models, demonstrating an average accuracy of 79\%.

The next set of experiments assesses the performance of technical CTI generation. In this experiment, we compared the most critical elements within the CTI reports, focusing on Indicators of Compromise (IoCs), Tactics, Techniques, and Procedures (TTPs), and Advanced Persistent Threats (APT). IoCs are data points that identify potential threats or breaches, including IP addresses, file hashes, and domain names associated with malicious activity. They are typically actionable and offer immediate insights into ongoing or past threats~\cite{preuveneers2021sharing}. TTPs represent the strategies and methodologies used by threat actors. This contextual information helps analysts understand not just \textit{what} happened but \textit{how} the attack was carried out, enabling more proactive defence measures~\cite{alam2023looking}. APTs are sophisticated, long-term cyber threat campaigns by highly skilled (typically state-sponsored) threat actors. Identifying APTs within a CTI report provides critical intelligence on the threat's origin, intent, and resources~\cite{wagner2019cyber}.
While IoCs provide direct evidence of compromise, TTPs offer a deeper understanding of the attack methodology. Each element serves a unique role in comprehensive CTI generation.

Table~\ref{tab:IoC} presents the results of technical CTI reports generated using AI, comparing them to their human-generated counterparts. The table's first and second columns (IoC and TTP) represent percentage performance accuracy. In contrast, the APT column indicates whether this element is included in the AI-generated reports. The designation \textit{N/A} signifies that the component is absent in both reports, likely due to its unavailability in the original artifacts. IoC and APT demonstrated the highest performance among the technical indicators, achieving 90.2\% and 85.7\% accuracy, respectively. In contrast, TTP showed relatively lower performance compared to the manually generated reports, with averages of 79.5\% accuracy.

\subsection{Automated CTI Report Generation Efficiency and Trade-offs}

While this research indicates the proof of concept, the system successfully automates the development of CTI reports, significantly speeding up the process and reducing the need for manual effort while maintaining the intelligence's accuracy. However, this efficiency comes with a cost—each report's creation requires substantial computational resources. These resources are primarily consumed by the AI tools employed, such as MCS and Azure AI Studio, which contribute to the environment's overall complexity and resource demands.

\textbf{Azure AI Studio:}  
The computational cost in Azure AI Studio is determined based on the Compute Instance (CI) specifications and the online endpoint where the prompt flow is deployed. The CI functions similarly to a Virtual Machine (VM) instance, with pricing in Azure based on the instance type and the number of hours it operates. All AI computations, such as prompt processing, are executed using the allocated computational resources, including the number of CPU cores, RAM, and storage space provided by the instance type. Each deployment functions like a web server, offering a URL endpoint that handles HTTP requests to trigger the prompt flow, returning the AI-generated output. Since the deployment endpoint is integrated into Logic Apps for automated HTTP requests, the compute instance must remain active 24/7. 
Azure AI Studio hosts two prompt flow deployments in this setup—one for processing Sections 2 and 3 of the report and another for Section 7. Each deployment is equipped with a 2-core CPU and 8GiB of RAM. The combined monthly cost for both deployments is approximately \$290, which breaks down to \$0.20 per hour per deployment (assuming a 30-day month with 720 hours).

\textbf{Microsoft Copilot for Security:} The usage cost for MCS is measured in Security Compute Units (SCUs), which are consumed during prompt processing and output generation. Two key factors influence the consumption of SCUs: the size of the prompt (the number of characters submitted to the model) and the utilization of specific \textit{skills} within MCS. 
This implementation optimizes prompt sizes to reduce SCU usage by summarizing explanations and eliminating unnecessary spaces and line breaks. While there is no precise measurement for SCU consumption based on the use of skills, observations during development indicate that skills accessing Microsoft Threat Defender tend to consume more SCUs. Since our system primarily relies on the campaign blog URL as the primary intelligence source, the skill that accesses web URLs (i.e., FetchURL) is used more frequently.
For a typical report generation involving 4 out of 7 prompts, SCU usage ranges between 3.0 and 3.5 SCUs. Given the cost per SCU of \$5.60, the total cost for generating a single report using MCS is approximately \$18.50 (\$5.60 x 3.3 SCU).
 Despite the variability in AI outputs, the overall costs remain relatively consistent. However, enterprises must also consider the additional cost of human intervention required to ensure report quality.

\begin{table}
	\begin{center}
		\caption{Comparing the key elements of manual and AI-generated reports, indicating whether each element is captured by the AI model. }\label{tab:IoC}
		\begin{tabular}{   
				p{0.2\textwidth}   
				>{\centering\arraybackslash}p{0.052\textwidth}   
				>{\centering\arraybackslash}p{0.052 \textwidth}  
				>{\centering\arraybackslash}p{0.05\textwidth}  
			} 
			\toprule
			Report & IoC\% & TTP\% & APT    \\
			\midrule
			\rowcolor[HTML]{EFEFEF}
			Attachment Exploit &100 &  54  & N/A   \\
			
			OneNote &  85 & 71 &   \checkmark 	\\
			
			\rowcolor[HTML]{EFEFEF}
			Phishing Campaign & 78 & 100 & \checkmark 	\\
			
			Phishing Attachment & 100 & 100 & \checkmark \\
			
			\rowcolor[HTML]{EFEFEF}
			Distribute  RATs &  87 & 87 & \xmark  	\\
			Quishing & N/A  &  75 &  \checkmark	\\
			
			\rowcolor[HTML]{EFEFEF}
			Social engineering & 100  & 71 &  \checkmark 	\\
			ArcaneDoor & 82   & 78 & \checkmark  	\\
			\midrule
			Average  & 90.2\%  &  79.5\% &   85.7\%\\
			\bottomrule
		\end{tabular}
	\end{center}
	\tiny
	\textit{N/A} indicating the element absence in both documents. 
\end{table}	

\section{Discussion and Challenges}

 The cost and SCU usage are approximate due to the inherent variability in each output generated by the AI, which differs in terms of length and word choice. Both Azure AI and MCS utilize OpenAI's GPT-4o model for prompt processing~\cite{Ayfie}. This model is designed to introduce an element of randomness, ensuring that identical prompts, input values, and parameters do not result in the same output on different occasions~\cite{AnkurJhaveriPost}. Despite this variability, the overall quality and intent of the generated content remain consistent.
 Due to the minor inconsistencies and the continuous advancements in AI technology, the outputs generated by the system cannot be considered final without human oversight. Human analysts must review and refine the AI-generated reports to meet the quality standards of manually produced reports. The primary goal of this research project was to accelerate the report creation process and reduce manual effort rather than completely eliminate the need for human involvement.
 
 To report the acceleration of the AI-based model, we interviewed the Threat Intelligence team at our industrial partner site. Based on the survey outcome,  the manually produced CTI report from raw data typically required an average of 8 hours of effort, depending on the threat's complexity. With the introduction of AI automation, the manual effort needed by an analyst has been reduced to approximately 1 to 2 hours. This remaining work primarily involves refining the AI-generated content by correcting minor inaccuracies, enhancing the language, and verifying the information. As a result, this approach increases report creation speed by reducing manual endeavours.
 The main limitations of AI-driven report generation stem from the inherent variability and minor inconsistencies in AI-generated outputs due to the randomness introduced by OpenAI's GPT-4o model. This variability requires human intervention to review and refine the reports, ensuring they meet the quality standards of manually produced ones. Furthermore, human oversight remains crucial for maintaining the accuracy and reliability of AI technology.

\section{Conclusion and Future Work}

This study presents an architecture for automating Cyber Threat Intelligence (CTI) report generation using AI-powered tools within the Microsoft ecosystem. The primary objective was to accelerate the CTI generation process, enabling faster sharing of intelligence and enhancing attack detection and prevention for stakeholders. This objective was successfully achieved by integrating Microsoft Copilot for Security, Azure DevOps, and Azure AI Studio. Our evaluation of the proposed architecture, comparing manually generated and AI-generated reports, demonstrated high performance across both technical and executive-level reporting. 
Notably, while AI performed equally well in generating executive summaries, human intervention is still crucial in the more complex technical sections of the report. This necessity underscores the value of human expertise in maintaining the level of precision necessary for effective decision-making and implementation by security teams.

Future research will focus on evaluating the proposed method with larger datasets, incorporating both manually and AI-generated reports. Additionally, an in-depth analysis of other threat types beyond campaign threat intelligence will be conducted to assess performance and other factors, including cost. While GPT-4o was the primary language model used for prompt engineering, future work will explore the use of alternative models to compare their effectiveness.
The experiments further indicated that combining AI-driven processes with human input yields the highest performance. As a next step, we propose developing a hybrid model that integrates other AI-based methods with proprietary security products. This model promises to further refine CTI report generation and enhance the overall quality of the final output, instilling optimism about the future of CTI report generation.

\section{Acknowledgement} 

We appreciate the collaborative efforts and support from 
Difenda 
throughout this project. This study was made possible by the resources provided by our industrial partner, for which we are grateful.
This research has been supported by 
 research fund 
FR124652.

\balance

\bibliographystyle{ieeetr}
\bibliography{reference}
%
%

\end{document}